\documentstyle [12pt,epsf]{article}
\begin{document}

\def\NPBsupl#1#2#3{Nucl. Phys. {\bf B} (Proc. Suppl.) {\bf #1} (#2) #3}
\def\NPB#1#2#3{Nucl. Phys. {\bf B #1} (#2) #3}
\def\PLB#1#2#3{Phys. Lett.  {\bf B #1} (#2) #3}
\def\PRD#1#2#3{Phys. Rev. {\bf D #1} (#2) #3}
\def\PRL#1#2#3{Phys. Rev. Lett. {\bf  #1} (#2) #3}
\def\PROG#1#2#3{Prog. Theor. Phys. {\bf #1} (#2) #3}
\def\CM#1#2#3{Comm. Math. Phys. {\bf #1} (#2) #3}
\def\PRL#1#2#3{Phys. Rev. Lett. {\bf  #1} (#2) #3}
\def\MPLA#1#2#3{Mod. Phys. Lett. {\bf A #1} (#2) #3}

\begin{titlepage}
\begin{flushright}
April 5 1993

UTHEP-256

\end{flushright}
\vskip 1.0cm
\center{\LARGE Charged fermion states in the quenched U(1) chiral
Wilson-Yukawa model}
\Large
\vskip 2.5cm
\center{S. Aoki,  H. Hirose and Y. Kikukawa}
\vskip 0.6cm
\center{Institute of Physics, University of Tsukuba,}
\center{Tsukuba, Ibaraki-305, Japan}
\vskip 3.0cm

\begin{abstract}

Property of charged fermion states is investigated
in the quenched U(1) chiral Wilson-Yukawa model.
Fitting the charged fermion propagator with a single hyperbolic cosine
does not yield reliable results.
On the other hand the behavior of the propagator
including large lattice size dependence is well described by
the large Wilson-Yukawa coupling expansion,
providing strong evidence that no charged fermion
state exists as an asymptotic
particle in this model.
\end{abstract}
\vfill
\end{titlepage}
\section{Introduction}

The effort toward construction of chiral gauge theories
on the lattice is primarily motivated by the fact
that a full nonperturbative understanding
of the standard SU(2) $\times$ U(1)$_Y$ electroweak theory
is required to answer a variety of unsolved questions such as
the upper bound of the Higgs boson and
fermion masses in the full theory
and the magnitude of the baryon number violation in the standard model.
The lattice approach provides a potentially useful tool for these purposes.
However, there is a serious difficulty in this approach:
because of fermion doubling\cite{NNtheorem,Karsten},
each lattice fermion field yields 16 fermion
modes, half of which has one chirality and
the other half the opposite chirality, so that the lattice
theory is non-chiral.
There are several attempts to avoid this difficulty.
Among the proposals, the Wilson-Yukawa formulation
\cite{Swift,Smit,Aoki,Kashiwa}
has been extensively investigated during the last few years.
In particular it has been found
through numerical studies\cite{BDJJNS,ALSS,BD} and
analytical work\cite{Mean,ALX,Hop,IvsD}
that the fermion doublers
decouple in the continuum limit
if the Wilson-Yukawa coupling is sufficiently large.

Recently it has been reported\cite{BDS}
in a quenched simulation of a $\rm U(1)$ chiral model
that the charged fermion mass
is approximately equal to the sum of the neutral fermion mass
and the scalar boson mass on a $12^3\times 16$ lattice.
This relation suggests that the charged state
is a scattering state of the neutral fermion
and the scalar boson, and is not a bound state.
Based on this interpretation it was claimed that
charged fermion states do not exist
in the symmetric phase of the global chiral model
for strong Wilson-Yukawa coupling,
and hence the Wilson-Yukawa formulation fails.

It should be noted, however, that a large finite size effect
is seen in the data of ref.\cite{BDS}
for charged fermion propagators
between $8^3\times 16$ and $12^3\times 16$ lattices and
that the relation among the masses does not hold on an
$8^3\times 16$ lattice.
Thus much further analysis of the model,
especially a systematic study of finite size effect,
is needed to see if the conclusion is tenable.
We have carried out
such an analysis and we report on the results in this article.

The organization of this article is as follows.
In Sect.2, we define the action of the $\rm U(1)$ chiral model we study
and consider several bosonic and fermionic field operators,
especially the two kinds of Dirac fermions, neutral and charged.
We measure two-point correlation functions of these operators
in the simulation whose parameters are also summarized.
In Sect.3 spectra in the broken phase are investigated.
We extract masses of neutral and charged fermions
as well as those of scalar particles.
In Sect.4 we present and discuss
the results of our Monte Carlo simulation
in the symmetric phase in detail.
Emphasis is placed on the fermionic charged state,
which has been argued not to exist as a single particle
in this phase\cite{BDS}.
Finite size effects for the charged fermion state
are examined by means of the large Wilson-Yukawa coupling expansion.
Our conclusions are summarized in Sect.5.

\section  {$\rm U(1)$ chiral model}

\subsection{The action}

We investigate a chiral U(1) Higgs-fermion system
in the Wilson-Yukawa formulation with the action given by
\begin{eqnarray}
S_{total} &=& S_{boson} + S_{free} + S_{yukawa}+ S_{wilson-yukawa}
\nonumber \\
	  &\equiv & S_{boson} + S_{fermion},  \\
S_{boson} &=& -2 \kappa \sum_{n,\mu} Re (g_n ^\dagger  g_{n+\mu}), \\
S_{free} &=& - {1 \over 2} \sum_{n,\mu} \left(\bar \psi_n  \gamma_\mu
\psi_{n+\mu} - \bar \psi_{n+\mu} \gamma_\mu  \psi_n \right), \\
S_{yukawa} &=& y \sum_n \bar \psi_n
\left( g_n ^\dagger P_L + g_n P_R \right)\psi_n , \\
S_{wilson-yukawa} &=& -{w \over 2}\sum_{n,\mu} \bar \psi_n P_L
\left( g_{n+\mu} ^\dagger \psi_{n+\mu } + g_{n-\mu} ^\dagger \psi_{n-\mu }
-2 g_n ^\dagger \psi_n \right) \nonumber \\
& & -{w \over 2}\sum_{n,\mu} \bar \psi_n P_R g_n \left(\psi_{n+\mu}+
\psi_{n-\mu} -2\psi_n \right).
\end{eqnarray}
Here $g_n$ is a non-linear U(1) scalar field satisfying
$g_n ^\dagger g_n = 1$, and $\kappa , y$ and $w$
are the hopping parameter of the boson,
the Yukawa coupling constant,
and the Wilson-Yukawa coupling constant respectively.
The phase transition between the broken phase and the symmetric phase
occurs at $\kappa_c = 0.149$ in the quenched theory.
The action $S_{total}$ is invariant under a global $\rm U(1)_L \times U(1)_R $
transformation given by
\begin{eqnarray}
 g_n & \rightarrow&  \Omega_L g_n \Omega_R ^\dagger ,
 \makebox[1cm]{ } \Omega_{L,R} \in {\rm U(1)_{L,R}}, \nonumber \\
\psi_n  &\rightarrow & (\Omega_L P_L + \Omega_R P_R)\psi_n , \nonumber\\
 \bar  \psi_n  &\rightarrow  &
\bar \psi_n (\Omega_L ^\dagger P_R + \Omega_R ^\dagger P_L). \nonumber
\end{eqnarray}
The scalar field $g$ and the {\it chiral} fermion fields $\psi_L$
and $\psi_R$ are assigned the charges
$(Q_L, Q_R) = (1, -1)$, $(1, 0)$ and $(0,1)$, respectively.

We consider two kinds of {\it Dirac} fermion fields in our study
defined by
\begin{eqnarray}
N_n =(g_n ^\dagger P_L + P_R) \psi_n ,\makebox[2em]{} \bar N_n =
\bar \psi_n(g_n P_L+P_R) ,
\end{eqnarray}
with the charge (0,1) and
\begin{eqnarray}
 C_n =(P_L+ g_n P_R) \psi_n ,\makebox[2em]{} \bar C_n = \bar \psi_n
(P_L+g_n ^\dagger P_R)
\end{eqnarray}
with the charge (1,0).
The fermion fields $N_n $ and $C_n $
are usually called ``neutral'' and ``charged''
in view of the eventual gauging of the global $\rm U(1)_L$ symmetry.
This distinction is not important since $N_n$ will be
the ``charged'' fermion if the $\rm U(1)_R$ symmetry is gauged.
The most important distinction between $N_n$ and $C_n$ is that
the Wilson-Yukawa coupling is reduced to an ordinary Wilson term and
the Yukawa coupling to a mass term for the neutral fermion $N_n$.
In fact one easily sees that
\begin{eqnarray}
S_{fermion} & = & -{1 \over 2} \sum_{n,\mu} \bar N_n  \gamma_\mu P_L
g_n ^\dagger \left( g_{n+\mu} N_{n+\mu} - g_{n-\mu } N_{n-\mu}\right)
\nonumber \\
&  & -{1 \over 2} \sum_{n,\mu} \bar N_n  \gamma_\mu P_R \left( N_{n+\mu}
- N_{n-\mu } \right)  \nonumber \\
& & +y \sum_n \bar N_n N_n \\
& & -{w \over 2}\sum_{n,\mu} \bar N_n \left( N_{n +\mu} + N_{n-\mu}
-2 N_n \right) . \nonumber
\end{eqnarray}
Furthermore, the right-handed component
of the neutral fermion has the shift symmetry\cite{Shift}.
Consequently, it is expected to appear as a massless particle at
the point $y = 0$.
It has been shown (\cite{BDJJNS}--\cite{IvsD}) that the physical fermion state
of the $N_n$ field
appears in the continuum limit while doublers of the $N_n$ are decoupled
when the Wilson-Yukawa coupling $w$ is large enough ($ w \sim O(1)$~).

In terms of the charged fermion, on the other hand, the action is written as
\begin{eqnarray}
S_{fermion} & = & -{1\over 2}\sum_{n,\mu}\bar C_n \gamma_\mu P_L(C_{n+\mu}
-C_{n-\mu}) \nonumber \\
& & -{1\over 2}\sum_{n,\mu}\bar C_n \gamma_\mu P_R g_n
\left(g_{n+\mu} ^\dagger C_{n+\mu}-g_{n-\mu } ^\dagger C_{n-\mu} \right)
\nonumber \\
& & +y\sum_n \bar C_n C_n \\
& & -{w\over 2}\sum_{n,\mu} \bar C_n g_n \left(g_{n+\mu} ^\dagger
C_{n +\mu}+g_{n-\mu} ^\dagger C_{n-\mu}-2g_n ^\dagger C_n \right). \nonumber
\end{eqnarray}
Since $ w \sim O(1)$ is needed for the doublers to decouple,
the field $C_n$ interacts strongly with the scalar field
through the Wilson-Yukawa coupling.
We investigate this strongly coupled system
through the Monte Carlo simulation.

\subsection{Observables}

In order to extract the properties of the scalar state of the model,
we consider three different correlation functions
at zero spatial momentum defined by
\begin{eqnarray}
 D_\sigma (t)&=&  \sum_{\vec n}
<\sigma_{(\vec 0,0)}\sigma_{(\vec n,t)}>_c , \nonumber \\
 D_\pi (t)   &=&  \sum_{\vec n} <\pi_{(\vec 0,0)} \pi_{(\vec n,t)} >_c , \\
 D_g (t)&=& \sum_{\vec n}
<g^\dagger_{(\vec 0,0)} g_{(\vec n,t)} >_c ,\nonumber
\end{eqnarray}
where $\sigma_n = {\rm Re} (g_n)$, $\pi_n= {\rm Im} (g_n)$
and $<\qquad  >_c $ means the connected part of correlation functions.

In the symmetric phase (at $\kappa < \kappa_c = 0.149$
in the quenched approximation),
only one massive scalar state is expected.
Hence making a fit of form
\begin{eqnarray}
 D_{\phi} (t) = A_{\phi} \cosh[E_{\phi} (t - {\rm T}/2)]
\label{eqn:scal-ansat}
\end{eqnarray}
with {\rm T} the temporal lattice size should yield consistent values
for the energy $E_{\phi}$ independent of the choice $\phi = g, \sigma, \pi$.

The parameters $A_\phi$ and $E_\phi$ are related to
the wave function renormalization constant $Z_\phi$
and the mass parameter $m_\phi$ in the momentum-space-propagator through
\begin{eqnarray}
Z_\phi&=& 2 A_\phi \sinh(E_\phi) \sinh(E_\phi {\rm T}/2) , \nonumber \\
m_\phi^2 &= &2 \cosh (E_\phi) -2 .
\label{eqn:scal-ansat-sigma}
\end{eqnarray}

\bigskip

In the broken phase (at $\kappa < \kappa_c $),
the scalar field develops a vacuum expectation value
and the symmetry $\rm U(1)_L \times U(1)_R$
is broken down to $\rm U(1) _{L=R}$.
For a finite volume, however, fluctuation of
the Nambu-Goldstone mode makes the expectation value
of the scalar field vanish in absence of an external field.
To avoid this problem, each scalar field configuration
is rotated globally so that
it satisfies $\sum_n \pi_n = 0$. Consequently $v\equiv < g > = < \sigma >$
and the massive $\sigma$ propagator can be fitted by the same form as before.
The form of the propagator for the Nambu-Goldstone boson is determined
by the chiral perturbation theory\cite{Leutwyler}
which gives
\begin{eqnarray}
 D_\pi (t) = A + B \left(({t \over {\rm T}}
- {1 \over 2})^2 -{1 \over 12} \right),
\label{eqn:scal-pi-bro}
\end{eqnarray}
where $Z_\pi = 4 B / {\rm T}$.
{}From the constraint $ \sum_t D_\pi (t) = 0 $
$A$ and $B$ are related by $ A = - B / 6 {\rm T}^2 $.
In our fit we take $A$ and $B$ as free parameters
and check this relation for fitted results.

\bigskip

For the neutral fermion,
a general expression for the propagator
follows from the shift symmetry\cite{Shift}:
\begin{eqnarray}
S_F ^{-1} (p) &=& i \sum_\mu \gamma_\mu \sin(p_\mu)({\cal A}^L _\mu (p) P_L
+ {\cal A}^R _\mu (p) P_R) + {\cal B}(p), \\
{\cal A}^R _\mu (p) &=&  1  \equiv  1/Z^R , \\
{\cal B}(p)&=& y + w \sum_\mu (1-\cos(p_\mu)) .
\end{eqnarray}

The form of the function ${\cal A}^L _\mu (p)$
can be evaluated with several analytical
methods\cite{ALX,Hop,IvsD}.
The results are identical in the leading order, giving
\begin{eqnarray}
{\cal A}^L _\mu (p) &=&  < g_{n+\mu} ^\dagger g_n > = z^2  \equiv 1 / Z^L.
\label{eqn:scal-z2}
\end{eqnarray}
In this approximation, the propagator
at the zero spatial momentum takes the form
\[
\sum_{\vec n}  <N_{(\vec 0,0)} \bar N_{(\vec n,t)} >
= P_L \gamma_0 S_N^L(t) + P_R \gamma_0 S_N^R(t) + S^M _N(t) .
\]
where
\begin{eqnarray}
S^{L,R} _N (t)
&=& A^{L,R} _N \left({\cosh(E^- _N (t-{\rm T}/2))
\over \cosh(E^- _N {\rm T}/2)}
- {\cosh(E^+ _N (t-{\rm T}/2))
\over \cosh(E^+ _N {\rm T}/2) }\right) ,\nonumber \\
S^M _N (t)
&=& A^M _N \left({\sinh(E^- _N (t-{\rm T}/2))
\over \cosh(E^- _N {\rm T}/2)}\cdot
{ m+r(1-\cosh E_N^-) \over \sinh E^-_N } \right.  \nonumber \\
& & \left. -{\sinh(E^+ _N (t-{\rm T}/2))
\over \cosh(E^+ _N {\rm T}/2) }\cdot
{ m+r(1-\cosh E_N^+) \over \sinh E^+-N }\right)
\label{eqn:neut-ansat}
\end{eqnarray}
with
\begin{eqnarray}
m &=& y \sqrt{ Z ^L Z ^R  }, \makebox[1em]{}  r = w  \sqrt{ Z ^L Z ^R  } ,\\
A _N ^{L,R} &=&  { Z^{L,R}  \over  2 (m^2+2rm+1) ^{1/2}}, \\
A _N ^M &=&  { \sqrt{Z^L Z^R}  \over  2 (m^2+2rm+1) ^{1/2}}, \\
\exp(E_N ^\pm ) &=& {m+r+ \sqrt{m^2+2rm+1} \over r \mp 1}.
\end{eqnarray}
Here $E^- _N$ ($E^+ _N$) is the mass of the physical
fermion (the time doubler).
With the replacement $y \rightarrow y + 2 w$,
the formula (\ref{eqn:neut-ansat}) applies
to the (spatial) doubler mode at $\vec p = (\pi,0,0)$.
We denote its mass as $E_{ND}$.

We use the form (\ref{eqn:neut-ansat}) taking $ E_N ^\pm$
and $A^{L,R,M}_N$ as free parameters
in order to fit $S_N ^L$, $S_N ^R$ and $S_N ^M$ separately.
The doubler propagator is fitted
with the first term in (\ref{eqn:neut-ansat})
because its ``time doubler'' is much heavier.
We will also compare data
to the semi-analytical prediction (\ref{eqn:neut-ansat})
with the Monte Carlo estimation of $z^2$ as an input.

\

For the charged fermion,
the form of the propagator is not known.
We found that the formula (\ref{eqn:neut-ansat})
does not fit the charged fermion propagator well.
We therefore tried a single pole form\cite{BDS}
for the fit given by
\begin{eqnarray}
 S_C^{L,R} (t) =  A_C ^{L,R} \ \cosh(E_C(t-{\rm T}/2)).
\label{eqn:char-ansat}
\end{eqnarray}
We extract the mass $E_C$
using the data over a time interval
$t_{min} \le t \le {\rm T} - t_{min} ,$
varing $t_{min}$ in order to
see whether the mass is stable.

\subsection{Simulation }

 Our simulation is performed in the quenched approximation
on $8^3 \times 16$ and $12^3 \times 16$ lattices.
Scalar field configurations are generated
by the 10 hit Metropolis algorithm
and the cluster algorithm of ref.\cite{Wolf}
at $\kappa = 0.16$ (broken phase) and at $\kappa = 0.145$ (symmetric phase).
See Table \ref{table:scal-sim} for statistics.

The fermion propagators are calculated by the conjugate gradient method
with the point-source in the broken phase
and the wall-source in the symmetric phase.
The anti-periodic (periodic) boundary condition
in the time direction (the spatial directions ) is used.
We fix the Wilson-Yukawa coupling at $w = 1$
and vary the Yukawa coupling $y$.
The statistics depends on the simulation parameters.
Typically, 2,400 (400) configurations separated by 500 (200) sweeps
are accumulated at $\kappa = 0.145 (0.16)$
for the charged fermion propagator
on an $8^3 \times 16$ lattice.
See Table \ref{table:fermi-sim} for further details.

Errors are all estimated with the jack-knife method,
using the bin size as listed in Table \ref{table:scal-sim}
and Table \ref{table:fermi-sim}.
For example, for bosonic observables
on an $8^3 \times 16$ lattice at $\kappa = 0.16$,
each bin consists of the average
of successive 20,000 (10,000) sweeps
and the number of the bins is 4 (8)
with the Metropolis (cluster) algorithm.

\newpage

\section{Results for the broken phase}

The $\pi$ propagator in the broken phase
is shown in Fig.\ref{figure :scal-prop-broken}
together with the fit of form (\ref{eqn:scal-pi-bro})
expected from the chiral perturbation theory.
Fit is excellent with $A=-0.0218(52)$ and $B=23.7(2)$, which satisfies the
constraint $A= -B/6{\rm T}^2=-0.0154(1)$ within $1.3\sigma$.

The $\sigma$ propagator is shown
in Fig.\ref{figure :scal-prop-broken-cos}
together with the fit of form (\ref{eqn:scal-ansat}).
The data points around $ t = {\rm T}/2$ lie above the fitting curve.
We observed a similar behavior for the scalar field propagator
obtained with the cluster algorithm\cite{Wolf}.
This behavior may be related to the rotation procedure for calculating
$D_{\sigma}(t)$ in the broken phase.

The masses and the wave function renormalization constants $Z_\phi$
as well as the values of $v$ and $z^2$
are listed in Table \ref{table:scal-mass}.
{}From the results we estimate the renormalized self coupling constant:
\begin{eqnarray}
\lambda_R &=& 3 ({a m_\sigma \over a v}) ^2 Z_\sigma = 32.4(2).
\label{eqn:lamd}
\end{eqnarray}
The tree-level unitarity bound for the O(n) scalar model is
$\lambda_R < 48{\rm \pi}(n+1)^{-1}=50.2 (n =2)$\cite{Luscher}.

\bigskip

In the broken phase we expect
the relation $E_N = E_C$ between the energy of
the charged and the neutral fermion
since they have the same quantum number
under the residual $\rm U(1)_{L+R}$ symmetry.
This also can be seen from the following equation
\[ < C_0 \bar C_n > = v^2 < N_0 \bar N_n > + \cdots .\]
We fit the neutral fermion propagator with (\ref{eqn:neut-ansat})
and the charged fermion propagator with (\ref{eqn:char-ansat}).
Our results for $E_N$ and $E_C$ are plotted
in Fig.\ref{figure:fermi-YE-bro} by open circles and squares.
(see Table \ref{table:fermi-mass-broken} for numerical values.)
They agree perfectly and also with the analytic prediction of the hopping
parameter expansion\cite{Hop,IvsD,GPIvsW} drawn by dotted lines.
Thus our results in the broken phase confirm
the relation $E_N = E_C$ assumed in refs.\cite{BDJJNS,BD}.
(The relation $E_N \simeq E_C$ in the broken phase has been also observed
in ref.\cite{ALSS} using the momentum space propagators.)
It is also seen in Fig.\ref{figure:fermi-YE-bro} that
the mass of the neutral fermion doubler $E_{ND}$ (solid circles)
remains $O(1)$ in the $y \rightarrow 0$ limit.

\section {Spectrum in the symmetric phase}

\subsection{Scalar field}

The simulation has been performed with the parameters summarized in
Table \ref{table:scal-sim}.
The scalar propagator $D_g(t)$ is
shown in Fig.\ref{figure:scal-prop-symm} together with the fit.
It is seen that finite-size effects are small.
The result for the mass of the scalar field is summarized
in Table \ref{table:scal-mass}.
We have checked that these values are insensitive to the fitting range.
Our value $E_g = 0.415(1)$  on an $8^3\times 16$ lattice
obtained with the Metropolis algorithm
is substantially larger than
the value\footnote{This is calculated from the value $m_g = 0.390(9)$
quoted in ref.\cite{GPIvsW} using(\ref{eqn:scal-ansat-sigma})}
$E_g=0.387(9)$ of ref.\cite{BDS},
which employed the cluster algorithm\cite{Wolf}.
In order to make a direct comparison
we generated 120,000 configurations with the same cluster algorithm and
found $E_g = 0.413(3)$ on an $8^3\times 16$ lattice, which perfectly agrees
with our Metropolis value quoted above.
Thus the discrepancy is not caused
by a different choice of the algorithm.
For those who may be interested in resolving this problem,
we provide our data of $D_g(t)$ in Table \ref{table:scal-prop-symm}.

\subsection{Neutral fermion}

The calculation of the neutral fermion propagator
has been performed with the parameters in
Table \ref{table:fermi-sim}.
The propagator data on an $8^3 \times 16$ lattice
are shown in Fig.\ref{figure:neut-prop-symm}.
The propagators calculated through (\ref{eqn:neut-ansat})(dotted curves
in Fig.\ref{figure:neut-prop-symm}) agree with the data.
Fitted values of the energy $E_N ^{\pm}$
are given in Table \ref{table:fermi-mass}
and are plotted as a function of $y$ in Fig.\ref{figure:neut-YE-symm}
together with the prediction of the hopping parameter expansion.
Again it is seen that the hopping parameter expansion
agrees with the data.
Finite size effects for $E_N$ are very small.

\subsection{Charged fermion}

The propagator of the charged fermion $S_C ^L$ is shown
in Fig.\ref{figure:char-BDS}.
We observe a large finite size effect
between $8^3 \times 16$ and $12^3 \times 16$ lattices.
As mentioned before
we found that the propagators can not be fitted
by the free fermion ansatz.
We have therefore tried to extract the mass through a fit of the form
(\ref{eqn:char-ansat})
over the range $ t_{min} \leq t \leq {\rm T} - t_{min}$.
The value of $E_C$ as a function of $t_{min}$ is plotted in
Fig.\ref{figure:char-stabl}.
Since the value of $E_C$ significantly changes with the variation of
$t_{min}$, the charged fermion mass
can not be extracted reliably by the fit.
We are therefore not able
to confirm the relation claimed in ref.\cite{BDS}
that $E_C \simeq E_N + E_g$ on a $12^3\times 16$ lattice.
We should comment that our propagator numerically differs from that of
ref.\cite{BDS} by 3 to 5 standard deviations,
exhibiting a faster decrease with $t$.
(See Fig.\ref{figure:char-BDS}.)
Our data of $S_C ^L$ for both sizes are
given in Table \ref{table:scal-prop-symm}.
The discrepancy probably originates from the difference of the scalar field
configurations noted before.

\subsection{Interpretation of charged fermion state}

In order to understand the nature of the charged fermion state,
we have to find a good explanation for both the $t$ dependence of
the rest energy $E_c$ and the large finite size effect of the propagator.

It has been reported in ref.\cite{GPIvsW} that the large
Wilson-Yukawa coupling expansion explains the data of ref.\cite{BDS}.
However, the agreement is only qualitative.
In addition our charged fermion propagator is different
from that of ref.\cite{BDS}
and their results for the scalar mass
and the wave function renormalization constant,
which were used in the calculation of ref.\cite{GPIvsW} as an input,
 are also different from ours.
Furthermore the large finite size effect
of the charged fermion propagator was
not discussed in terms of the large Wilson-Yukawa coupling expansion
in ref.\cite{GPIvsW}.
We therefore think that a reanalysis of our data
in terms of the large Wilson-Yukawa coupling expansion is neccesary.

We have performed the expansion
for the charged fermion propagator
up to the next-to-leading order as in ref.\cite{GPIvsW}.
To this order the analytic expression for $S_C ^L (t)$ is given by
\begin{eqnarray}
S_C ^L (t) &= & {\alpha \over 4}^2 S_{conv.} ^L (t)  \nonumber \\
&-& { 1\over 2}({\alpha \over 4})^4 \left(
\sum_{t_1} S_{conv.} ^L (t-t_1) (\nabla_{t_1} S_{conv.} ^L (t_1))
+ W_L ^{(4)}(t) \right) ,
\label{eqn:1w}
\end{eqnarray}
where $\alpha = 4/(4 w + y) $ is the expansion parameter and
$\nabla_t f(t) = f(t+1) - f(t-1) $.
The leading order contribution
\begin{eqnarray}
S_{conv.} ^L (t) &=&  {1 \over {\rm L}^3}
\sum_{\vec k} S_N ^L (t, \vec k) D_g(t,-\vec k)  \nonumber
\end{eqnarray}
is the convolution of the neutral fermion propagator
\begin{eqnarray}
S_N ^L (t, \vec k) &=& {1 \over {\rm T} } \sum_{k_0}
{\rm e} ^{i k_0  t } {-i\sin(k_0) \over (\alpha z / 4 )^2
  \sum_\lambda \sin ^2 (k_\lambda) + (1-\alpha w  /4
\sum_\lambda \cos (k_\lambda))^2 } \nonumber
\end{eqnarray}
and the scalar propagator
\begin{eqnarray}
D_g (t, \vec p) &=& {1 \over {\rm T }} \sum_{p_0}
{\rm e} ^{i p_0  t }
{ Z_g \over 2 \sum_\lambda (1- \cos (p_\lambda)) + m_g ^2}
 \  . \nonumber
\end{eqnarray}
The boundary conditions in the time direction
imply $k_0 {\rm T} = (2n+1)\pi$ for the fermion field
and $p_0 {\rm T} = 2n \pi$ for the scalar field,
where $n=0,1,\cdots, {\rm T -1 }$.
In (\ref{eqn:1w}), $W_L ^{(4)}(t)$ is the left-handed part
of the 4-point-vertex contribution given by
\begin{eqnarray}
W^{(4)}(n_0) &=&  \sum_{\vec n } \sum_{\xi ,\mu}
\left( S_N (n-\xi) \gamma_\mu S_N(\xi+\hat \mu)
- S_N (n-\xi - \hat \mu) \gamma_\mu S_N (\xi) \right) \nonumber \\
&& \qquad \qquad \qquad
\times G^{(4)} (n , \xi, \xi + \hat \mu ) ,
\nonumber
\end{eqnarray}
with
\begin{eqnarray}
 S_N (n) &=& (P_R + P_L {4 \over \alpha})
<N_{(\vec 0,0 )}N_{(\vec n ,n_0 )}>(P_L + P_R {4 \over \alpha}),
\nonumber
\end{eqnarray}
where $G^{(4)}$	is the bosonic 4-point Green's function.
Following ref.\cite{GPIvsW} we take the mean-field result for $G ^{(4)}$.
Similar formulae can be obtained for $S_C ^R$ and $S_C ^M$.
See ref.\cite{GPIvsW} for details.

In order to compute the right handed side of (\ref{eqn:1w})
we use the values of $Z_g$, $m_g$ and $z^2$
determined in our numerical simulation results for $S_C ^L$
and the simulation data for the two lattices
are compared in Fig.\ref{figure:char-1w} for $y = 0.4$
and in Fig.\ref{figure:char-1w-0.1} for $y = 0.1$.
The agreement between the data and the prediction
is excellent at $y = 0.4$
and is very good even at $y = 0.1$.

The leading order contribution to $S_C ^L$
is the convolution of the neutral fermion propagator
and the scalar propagator,
each of which has very small lattice size dependence.
It turns out, however,
that their convolution has a large size dependence,
which explains the size effect in our data.
The next-to-leading order correction is found to be relatively small.
The analytic predictions for $S_C ^R$ and $S_C ^M$ also
agree with the data very well.
(See Fig.\ref{figure:char-1w-SR} and Fig.\ref{figure:char-1w-SM}).

\

As a further check,
we have performed the large Wilson-Yukawa coupling expansion
of the charged fermion propagator in the broken phase.
In this case the disconnected part has to be separated out from
 the two-point function of the scalar field,
and the connected part is the sum of contributions
from the massless Nambu-Goldstone boson
and the massive real scalar particle.
Therefore, we should add the following contribution
\[v^2 \sum_{\vec k} S_N ^L (t, \vec k)  \]
to (\ref{eqn:1w}) and replace the scalar propagator $D_g (t, \vec p)$
with the sum of those for the Nambu-Goldstone boson
and for the real scalar particle:
\[ {1 \over {\rm T }}{\sum_{p_0}^{ } }^\prime {\rm e} ^{i p_0  t }
{ Z_\pi \over 2 \sum_\lambda (1-\cos (p_\lambda)) }
+{1 \over {\rm T }} \sum _{p_0} {\rm e} ^{i p_0  t }
{ Z_\sigma \over 2 \sum_\lambda
(1-\cos (p_\lambda)) + m_\sigma ^2}.\]
Here $\sum ^\prime $ means
that $p_0 = 0$ is omitted in the sum for $\vec p = 0$.
The results up to the next-to-leading order
are plotted in Fig.\ref{figure:char-1w-bro}.
The expansion works well also in the broken phase.

Within the large Wilson-Yukawa coupling expansion
it has been shown in ref.\cite{GPIvsW} that
the charged fermion is a scattering state of the scalar particle
and the neutral fermion.
The good agreement between the data and the result of the expansion
therefore provides strong evidence that no charged fermion exists
as a single particle state in the symmetric phase.

\section{Conclusion}

We have investigated the spectrum of fermions
in the Wilson-Yukawa formulation
through a quenched simulation of the U(1) chiral Wilson-Yukawa model.
In the broken phase
the neutral fermion state exists
and the charged fermion operator generates the same state.
In the symmetric phase
the charged fermion mass can not be extracted
by a single hyperbolic cosine
fit even at large $t$ and charged fermion propagators suffer
a large finite size effect. We found that not only this large finite
size effect but also the form of charged fermion propagators
can be explained by the large Wilson-Yukawa coupling expansion.
Within the expansion the charged fermion in the symmetric phase
is interpreted as a scattering state
of the scalar boson and the neutral fermion.
We conclude that no charged fermion exists
as a single particle asymptotic state in the Wilson-Yukawa
model in {\it both} phases.
Only the scalar boson and the ``neutral'' Dirac fermion
exist in this model.
Although our data differ
from those of ref.\cite{BDS} and our reasoning
which leads to our conclusion is quite different,
we agree on the nature of the charged fermion state
in the Wilson-Yukawa formulation.

We can generalize this conclusion as follows. In more complicated chiral
models with the Wilson-Yukawa coupling,
we can always redefine fermionic field variables
so that the Wilson-Yukawa coupling
turns into a free Wilson mass term.
Only those states generated by such field variables
can appear as fermionic single particle states
in the strong Wilson-Yukawa coupling region.
And these states are necessarily
Dirac fermions with vector-like global charges.
Their interaction with the scalar fields are weak in general.

\section*{Acknowledgments}

Numerical calculations for the present work were carried out on HITAC
S820/80 at KEK.
We thank the Theory Division of KEK for warm hospitality.
The data analysis program SALS\cite{Oya}
was used for fittings in this article.
We also thank Prof. Kanaya and Prof. Ukawa for valuable discussions.
This work is supported in part by the Grant-in-Aid of the Ministry
of Education (No. 04NP0601).
Y.K. is supported by a Grant-in-Aid for JSPS fellow.

\newpage

\center{\bf Table captions}

\begin{description}
\item[Table 1.] Simulation parameters for scalar fields.
The symbol K represents 1,000 configurations.
\item[Table 2.] Simulation parameters for fermions.
Metro. means Metropolis algorithm.
Separation means the number of Monte Carlo sweeps between configurations.
\item[Table 3.] Parameter $E_{\phi}$ and
wave function renormalization constant $Z_{\phi}$ for scalar fields
and the value of $v$ and $z^2$.
\item[Table 4.] Fermion masses in the broken phase
on an $8^3\times 16$ lattice.
The minimum value of $t$ for fit is denoted by $t_{min}$.
\item[Table 5.] Numerical value of
the scalar field propagator $D_g(t)$ in the symmetric phase
at $\kappa = 0.145$ on an $8^3\times 16$ lattice.
\item[Table 6.] Neutral fermion mass in the symmetric phase.
\end{description}

\newpage
\center{\bf Figure captions}

\begin{description}
\item[Figure 1.] $\sigma$ propagator $D_\sigma (t)$
in the broken phase on an $8^3\times 16$ lattice.
The solid line represents the fit.
\item[Figure 2.] Same as Fig.1 for the $\pi$ propagator $D_\pi (t)$.
\item[Figure 3.] Fermion masses as a function of $y$
in the broken phase on an $8^3\times 16$ lattice.
Open circles (filled circles) represent the neutral fermion (neutral doubler)
mass and open squares represent the charged fermion mass.
The dotted line represents the prediction
of the hopping parameter expansion (HPE) for the neutral fermion
mass.
\item[Figure 4.] Propagator $D_g (t)$ for the scalar field $g$
in the symmetric phase on an $8^3\times 16$ lattice
(filled circle) and a $12^3\times 16$ lattice(open squares).
The solid line represents the fit for the data on a $8^3\times 16$ lattice.
\item[Figure 5.] Neutral fermion propagator $S_N^L (t)$
at $y=0.4$ (filled circles) and $y=0.1$ (open circles)
in the symmetric phase on an $8^3 \times 16$ lattice.
Dotted lines represent the prediction of the hopping parameter expansion
(HPE) for neutral fermion propagators.
\item[Figure 6.] Neutral fermion masses as a function of $y$
in the symmetric phase on an $8^3 \times 16$ lattice
(filled circles) and a $12^3\times 16$ lattice(open squares).
Dotted line represents the prediction
of the hopping parameter expansion (HPE) for the neutral fermion mass.
\item[Figure 7.] Charged fermion propagator $S_C^L (t)$
at $y=0.4$ in the symmetric phase on an $8^3 \times 16$ lattice
(open circles) and a $12^3\times 16$ lattice (open squares).
Vertical bars connected by solid (dotted) lines
represent the data of ref.\cite{BDS} for the same parameters
on an $8^3\times 16$ (a $12^3\times 16$) lattice.
\item[Figure 8.] Fitted value of $E_C$ as a function of $t_{min}$ in
the symmetric phase at $y=0.4$ on an $8^3\times 16$ lattice (open circle)
and $12^3\times 16$ lattice (filled circle).
\item[Figure 9.] Charged fermion propagator $S_C^L (t)$
at $y=0.4$ in the symmetric phase
on an $8^3\times 16$ lattice (filled circle) and $12^3\times 16$ lattice
(open circle), together with the prediction of the large $w$ expansion
including the next-to-leading order(solid lines).
Dotted lines represent
the convolution of $S_N^L(t)$ and $D_g(t)$, which is the leading order
contribution in the expansion.
\item[Figure 10.] Same as Fig.9 at $y=0.1$.
\item[Figure 11.] Charged fermion propagators $S_C^R (t)$
at $y=0.4$ in the symmetric phase
on an $8^3 \times 16$ lattice (filled circle) and a $12^3\times 16$ lattice
(open circle), together with the prediction by the large $w$ expansion
(solid lines).
\item[Figure 12.] Same as Fig.11 for $S_C^M(t)$.
\item[Figure 13.] Charged fermion propagators, $S_C^L (t)$,
at $y=0.3$ in the broken phase.
The meaning of symbols are the same as the Fig.9.

\end{description}

\newpage
\pagestyle{empty}
\setlength{\unitlength}{1pt}

\begin{table}[h]
\caption [The parameter of the simulation (the scalar field)]{ }
\label{table:scal-sim}
\begin{center}
\begin{tabular}{|l||c|c||c|c|c|}
\hline \hline
& \multicolumn{2}{c||}{Broken } & \multicolumn{3}{c|}{Symmetric} \\
\hline
$\kappa$ & \multicolumn{2}{c||}{0.16} & \multicolumn{3}{c|}{ 0.145}  \\
\hline
size  &\multicolumn{2}{c||}{ $8 ^3 \times 16$} &
\multicolumn{2}{c|}{$8^3 \times 16$} & $12^3 \times 16$ \\
\hline
algorithm & Metropolis & Cluster & Metropolis & Cluster & Metropolis\\
\hline
\#sweeps & 80K & 80K  & 1,200K & 240K  & 900K \\
\hline
\#bin   & 4 & 8 & 12&24  & 18  \\
\hline
bin size  & 20K & 10K & 100K&10K  & 50K  \\
\hline \hline
\end{tabular}
\end{center}
\caption [The parameter of the simulation (fermion)]{ }
\label{table:fermi-sim}
\begin{center}
\begin{tabular}{|l||c|c||c|c|c|c|}
\hline \hline
 & \multicolumn{2}{c||}{Broken} & \multicolumn{4}{c|}{Symmetric} \\
\hline
$\kappa$  &\multicolumn{2}{c||}{0.16 (Metro.)}
& \multicolumn{4}{c|}{0.145(Metro.)} \\
\hline
size  & \multicolumn{2}{c||}{$8 ^3 \times 16$} &
\multicolumn{2}{c|}{$8^3 \times 16$} &
\multicolumn{2}{c|}{$12^3 \times 16$} \\
\hline
fermion & neutral & charged & neutral & charged & neutral & charged \\
\hline
y  &\multicolumn{2}{c||}{0.1,0.2,0.3} &0.1/0.4
& 0.1,0.4 &\multicolumn{2}{c|}{0.4}  \\
\hline
separation & \multicolumn{2}{c||}{200} & \multicolumn{2}{c|}{500} &
\multicolumn{2}{c|}{250} \\
\hline
\# conf. & \multicolumn{2}{c||}{400} &600/2,400 & 2,400 & 800 & 3,600 \\
\hline
\# bin & \multicolumn{2}{c||}{4} & 3 / 12  & 12 & 4 & 18 \\
\hline
bin size & \multicolumn{2}{c||}{100} & \multicolumn{2}{c|}{200} &
\multicolumn{2}{c|}{200} \\
\hline
source & \multicolumn{2}{c||}{point} & \multicolumn{4}{c|}{wall} \\
\hline \hline
\end{tabular}
\end{center}
\end{table}
\newpage
\begin{table}[h]
\caption[Mass and Z of the scalar fields]{ }
\label{table:scal-mass}
\begin{center}
\begin{tabular}{|l||c|c|c|c|c|}
\hline \hline
$\kappa$ & \multicolumn{2}{c|}{0.16}  & \multicolumn{3}{c|}{0.145} \\
\hline
size &\multicolumn{2}{c|}{$8^3 \times 16$} &
\multicolumn{2}{c|}{$8^3 \times 16 $} & $12^3 \times 16 $  \\
\hline
algorithm  & Metro.  & Cluster & Metro. & Cluster & Metro. \\
\hline
$E_\sigma$      & 0.809(3) & 0.813(30) & 0.415(2)& 0.419(4)   &  0.412(2) \\
$Z_\sigma$      & 2.78(15) & 2.76(18) & 3.36(13)& 3.33(19)   &   3.37(13) \\
\hline
$E_\pi$      & 0       & 0 &      0.416(3)& 0.409(4)   &    0.410(2)  \\
$Z_\pi$      & 5.94(5) & 5.87(8)  &  3.39(17)& 3.34(24)   &    3.37(12)  \\
\hline
$E_g$           & --  & --  &  0.415(1)& 0.414(3)    &    0.411(2)   \\
$Z_g$           & --  & --   &  6.73(13)& 6.66(23)   &    6.75(15)    \\
\hline
\hline
$v$             & 0.401(1) &0.402(1) &  --      &  --   &   --       \\
$z^2$           & 0.303(1) &0.304(1) & 0.177(1) &  0.177(4)   &    0.176(1) \\
\hline \hline
\end{tabular}
\end{center}
\end{table}
\begin{table}[h]
\caption[Fermion masses in the broken phase ($8^3 \times 16$)]{ }
\label{table:fermi-mass-broken}
\begin{center}
\begin{tabular}{|l|c||ccc|}
\hline \hline
$y$ & $t_{min.}$   &   0.1            &      0.2           &   0.3    \\
\hline
$E^{-} _N$ & 0   &     0.15(1)      &      0.277(7)      &   0.384(7) \\
$E^{+} _N$ & 0   &     1.7(2)       &      1.76(9)       &   1.71(8)  \\
$E_{ND}$   & 6   &      1.4(3)      &      1.3(2)        &   1.5(2)  \\
\hline
$E_C$ & 3 &0.149(24) &0.277(1)   &0.386(16)  \\
      & 4 &0.152(49) &0.276(11)  &0.38(2)   \\
      & 5 &0.168(83) &0.283(19)  &0.38(7)  \\
\hline \hline
\end{tabular}
\end{center}
\end{table}
\newpage
\begin{table}[h]
\caption[The propagator $D_g(t)$ in the symmetric phase ($8^3 \times 16$).]{ }
\label{table:scal-prop-symm}
\begin{center}
\begin{tabular}{|l||l|l|l|}
\hline \hline
 $t $ &  $D_g$ & \multicolumn{2}{c|}{$S_C ^L$} \\
\hline
      &  $8^3 \times 16$  & $8^3 \times 16$ & $12^3 \times 16$ \\
\hline
0 &7.905(9)& 0.083(98) ($\times 10^{-4}$) &   0.048(49) ($\times 10^{-4}$) \\
1 &5.227(9)& 0.1694(1) &   158.6(3) \\
2 &3.466(1)& 41.7(4)  &  33.1(3)\\
3 &2.31(1) & 13.6(5)  &   9.3(2)\\
4 &1.55(1) & 4.8(3)   &   2.6(7)\\
5 &1.07(1) & 2.2(1)   &   0.88(5)\\
6 &7.75(1) & 0.90(1)   &  0.33(3)\\
7 &6.16(1) & 0.49(6)   &  0.15(2)\\
8 &5.66(2) & 0.36(8)   &  0.12(1)\\
9  &6.16(1)& 0.47(9)&     0.17(2)\\
10 &7.75(1)& 0.85(1)&    0.35(3)\\
11 &1.07(1)&  1.9(1)&    0.87(5)\\
12 &1.55(1)&  4.8(3)&    2.8(1)\\
13 &2.31(1)&  13.1(5)&   9.3(2)\\
14 &3.466(1)& 40.2(6)&   33.1(3)\\
15 &5.227(9)& 168.8(1)&  159.9(5)\\
\hline \hline
\end{tabular}
\end{center}
\end{table}
\begin{table}[h]
\caption[Neutral fermion mass in the symmetric phase.]{ }
\label{table:fermi-mass}
\begin{center}
\begin{tabular}{|ll||cc|}
\hline \hline
size            &              & $y$ = 0.1        & 0.4             \\
\hline
$8^3 \times 16$ & $E_N ^{-}$ &  0.1895(1)       & 0.5491(6)     \\
                & $E_N ^{+}$ &  1.109(1)        & 1.458(2)         \\
\hline
$12^3 \times 16$& $E_N ^{-} $        & --              & 0.5495(1)    \\
                & $E_N ^{+}$         &  --             &   1.4562(9)   \\
\hline \hline
\end{tabular}
\end{center}
\end{table}
\newpage
\begin{figure}[h]
\begin{center}
\begin{picture}(360,400)
\epsfxsize=13.0cm
\epsfysize=7.0cm
\put(0,190){\epsfbox{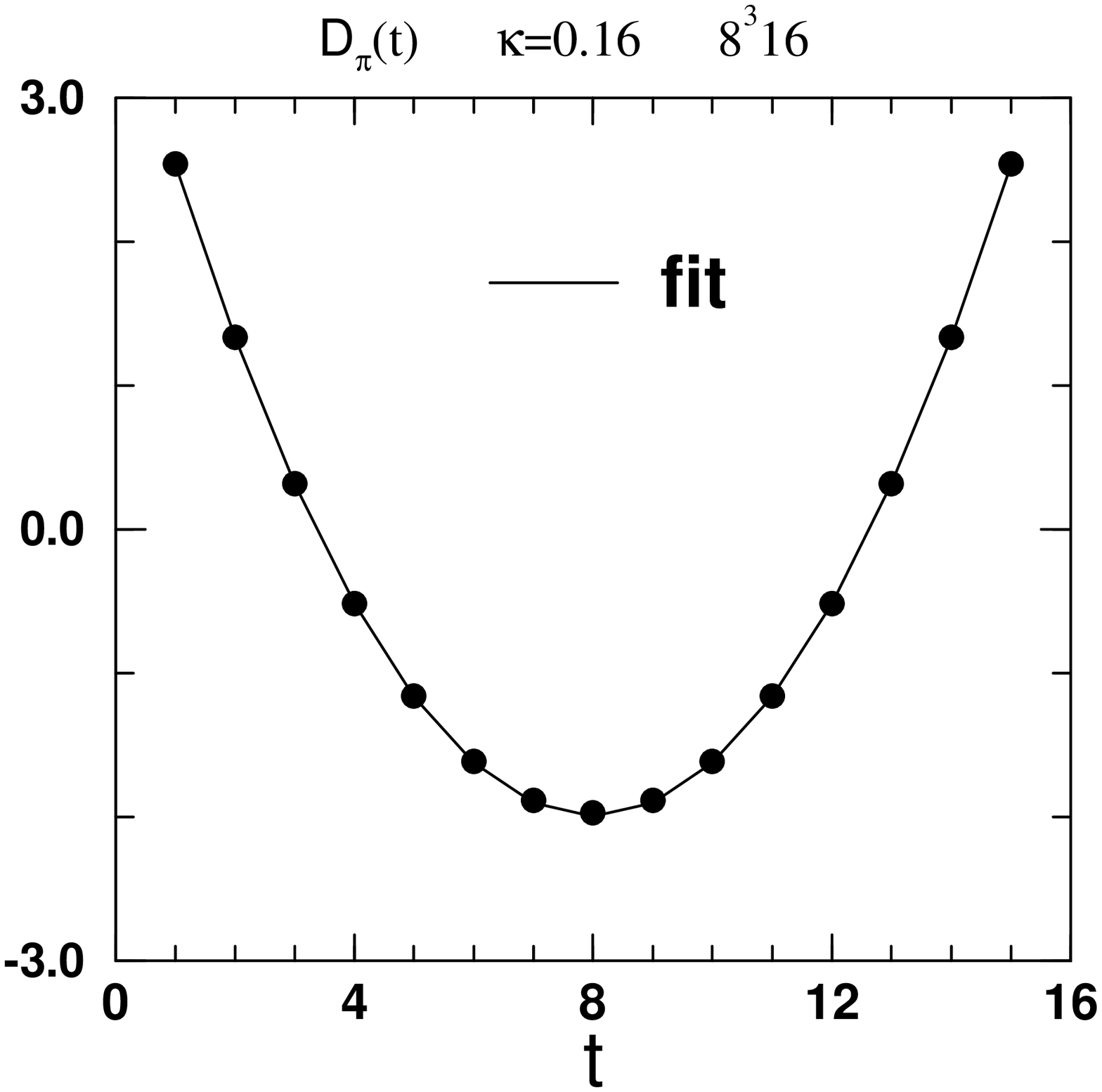}}
\end{picture}
\end{center}
\caption[The propagators of the $\pi$ in the broken phase ]{ }
\label{figure :scal-prop-broken}
\end{figure}
\begin{figure}[h]
\begin{center}
\begin{picture}(360,400)
\epsfxsize=13.0cm
\epsfysize=7.0cm
\put(0,190){\epsfbox{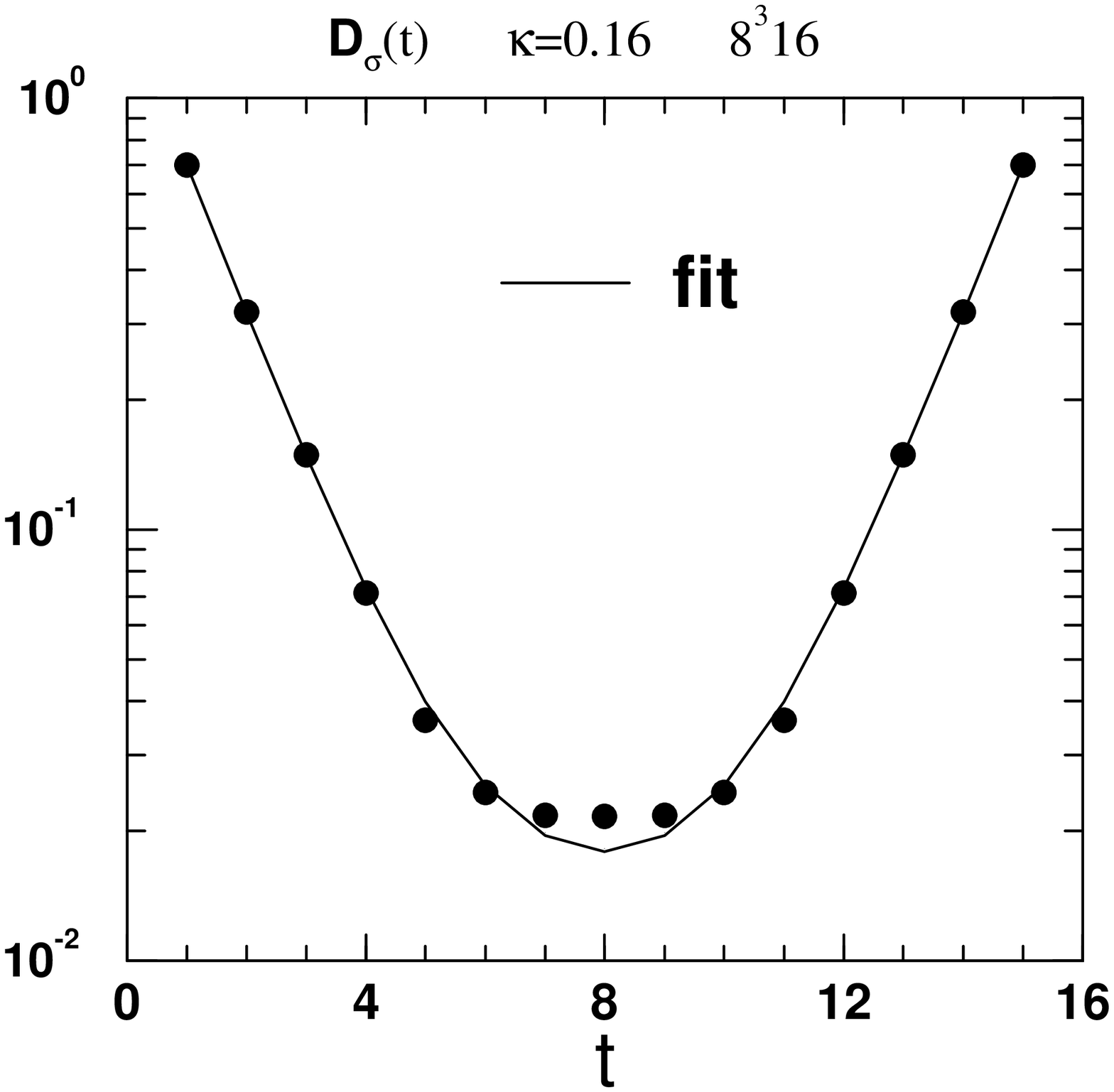}}
\end{picture}
\end{center}
\caption[The propagators of the $\sigma$ in the broken phase ]{ }
\label{figure :scal-prop-broken-cos}
\end{figure}
\begin{figure}[h]
\begin{center}
\begin{picture}(360,450)
\epsfxsize=12.0cm
\epsfysize=7.0cm
\put(0,190){\epsfbox{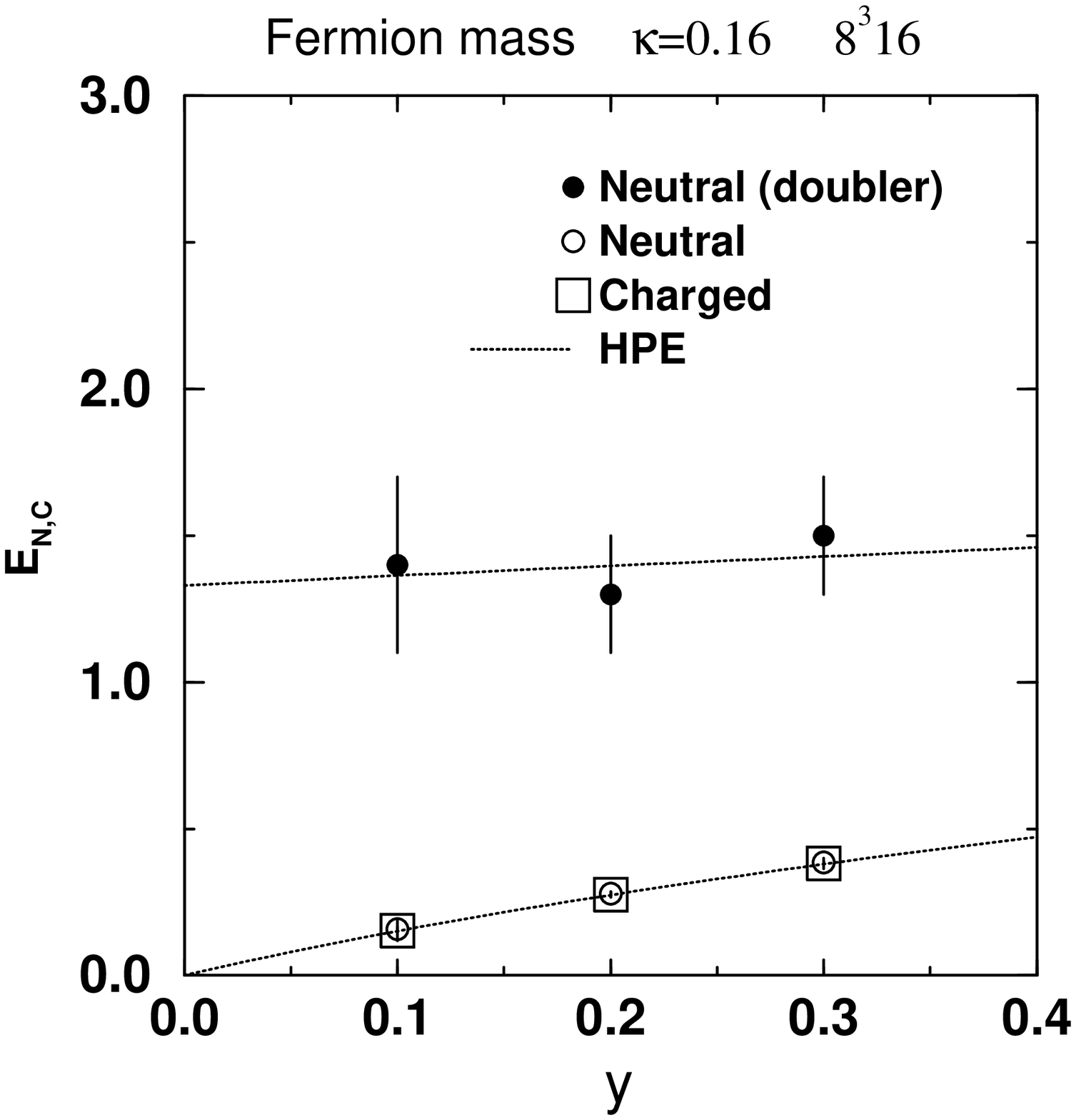}}
\end{picture}
\end{center}
\caption[The fermion mass vs $y$ in the broken phase ]{ }
\label{figure:fermi-YE-bro}
\end{figure}
\begin{figure}[h]
\begin{center}
\begin{picture}(360,450)
\epsfxsize=12.0cm
\epsfysize=7.0cm
\put(0,190){\epsfbox{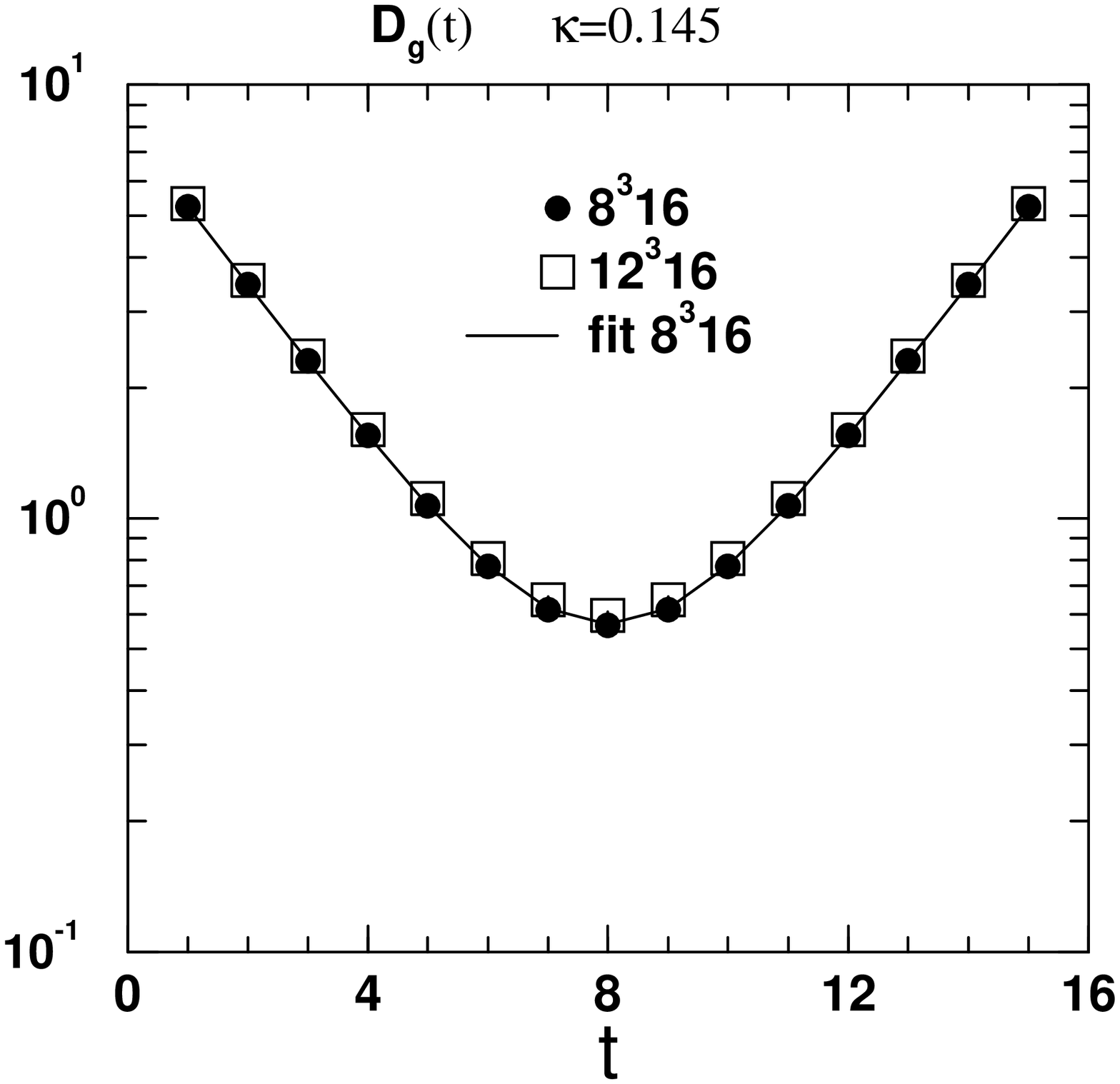}}
\end{picture}
\end{center}
\caption[The propagators of the scalar field
( $g$ ) in the symmetric phase ]{ }
\label{figure:scal-prop-symm}
\end{figure}
\begin{figure}[h]
\begin{center}
\begin{picture}(360,450)
\epsfxsize=12.0cm
\epsfysize=7.0cm
\put(0,190){\epsfbox{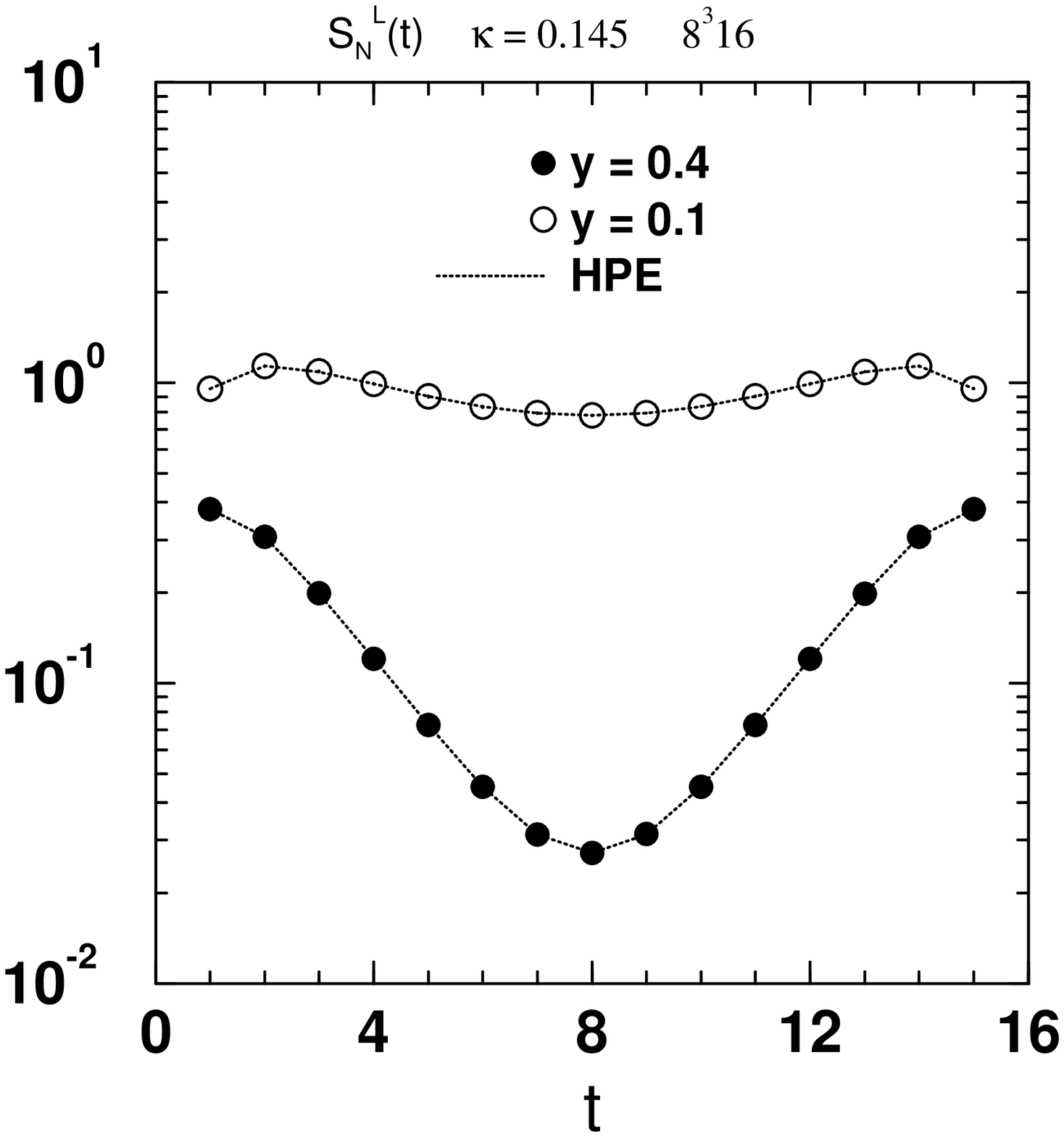}}
\end{picture}
\end{center}
\caption[The propagators of the neutral fermion ($S^L$)
in the symmetric phase at $ y = 0.1, 0.4$ on $8^3 \times 16$.]{ }
\label{figure:neut-prop-symm}
\end{figure}
\begin{figure}[h]
\begin{center}
\begin{picture}(360,450)
\epsfxsize=12.0cm
\epsfysize=7.0cm
\put(0,190){\epsfbox{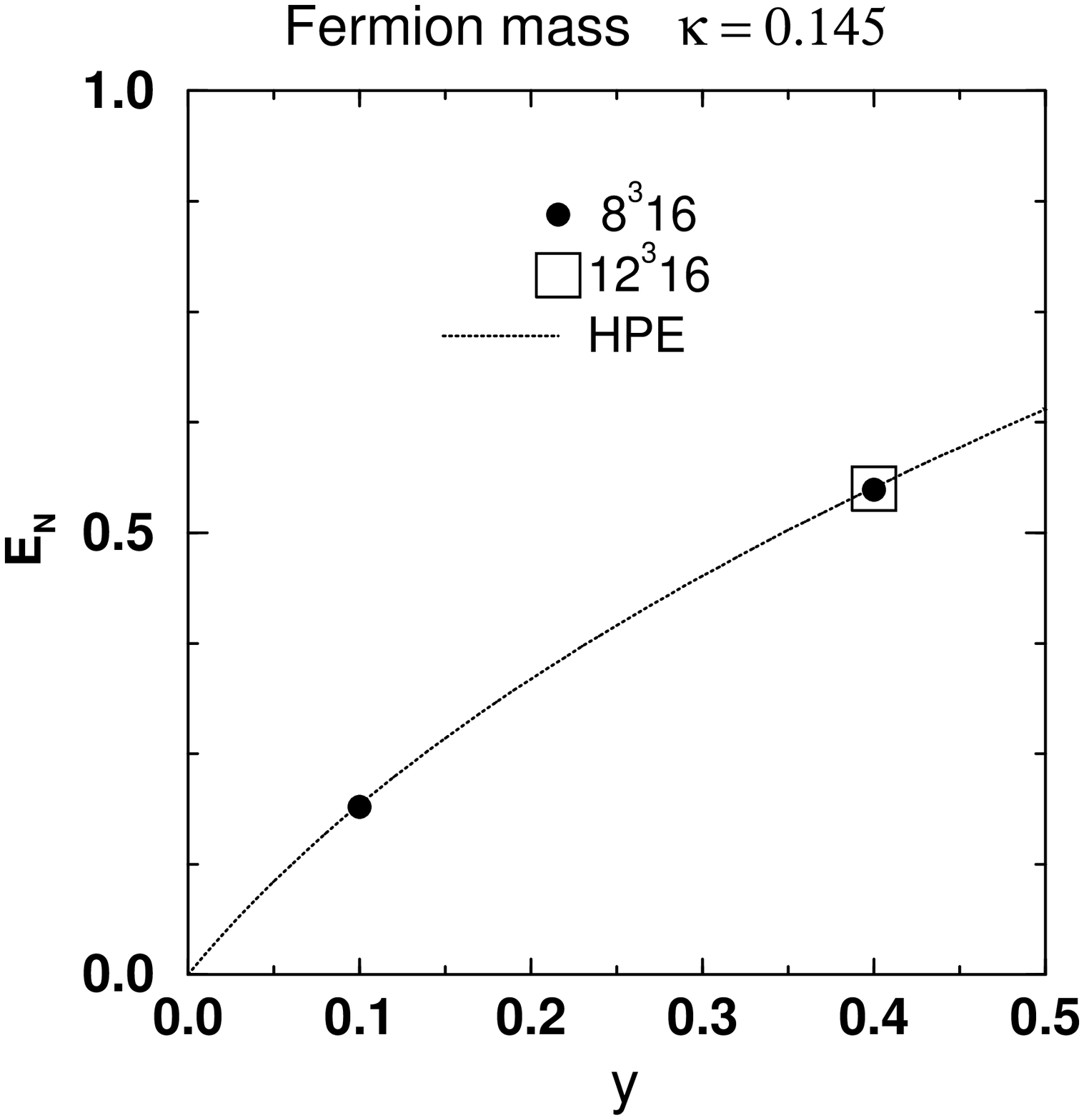}}
\end{picture}
\end{center}
\caption[The neutral fermion mass vs $y$ in the symmetric phase.]{ }
\label{figure:neut-YE-symm}
\end{figure}
\begin{figure}[h]
\begin{center}
\begin{picture}(360,450)
\epsfxsize=12.0cm
\epsfysize=7.0cm
\put(0,190){\epsfbox{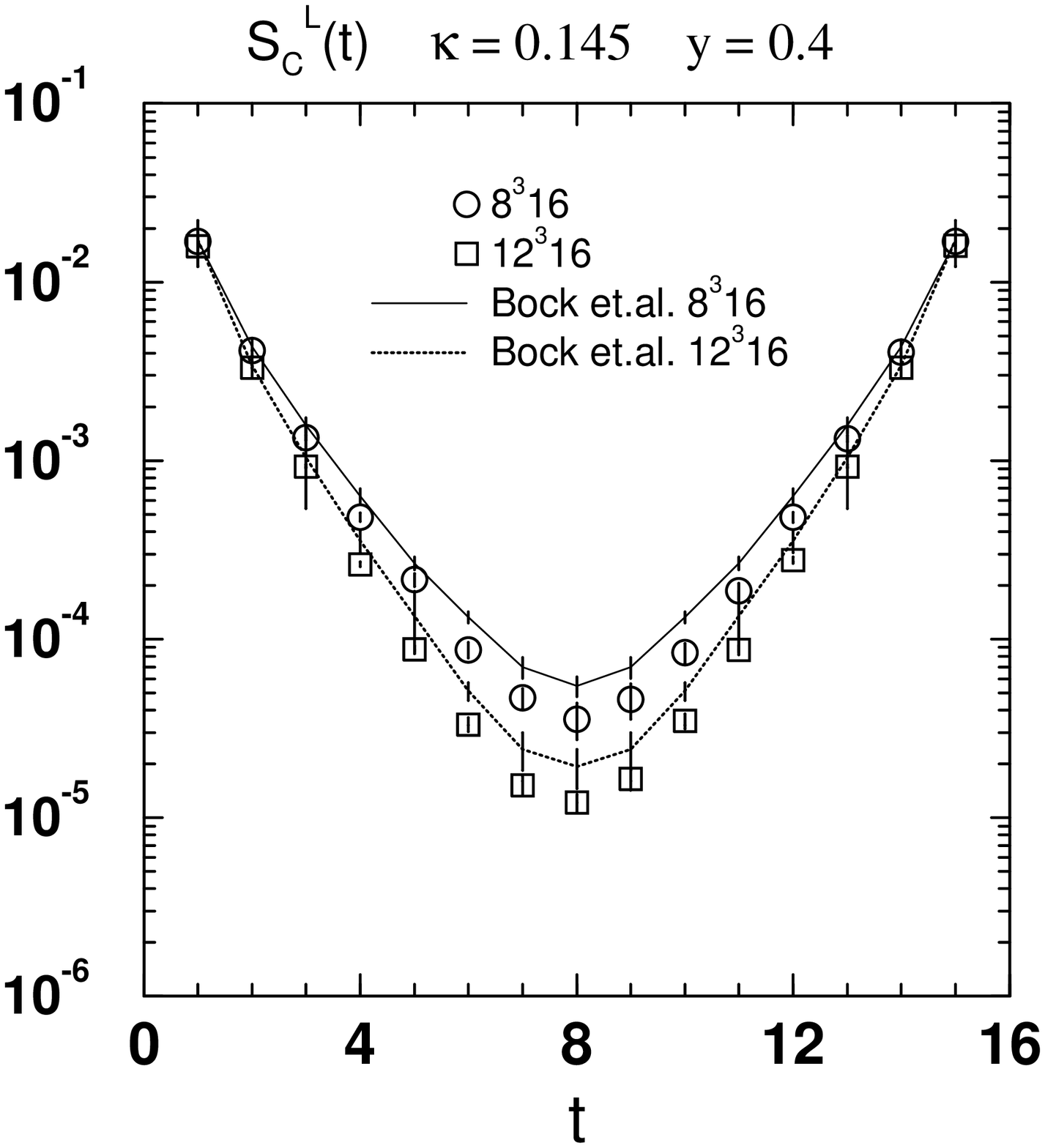}}
\end{picture}
\end{center}
\caption[The comparsion between BDS and our charged fermion propagator
$S^L$ in the symmetric phase.]{ }
\label{figure:char-BDS}
\end{figure}
\begin{figure}[h]
\begin{center}
\begin{picture}(360,450)
\epsfxsize=12.0cm
\epsfysize=7.0cm
\put(0,190){\epsfbox{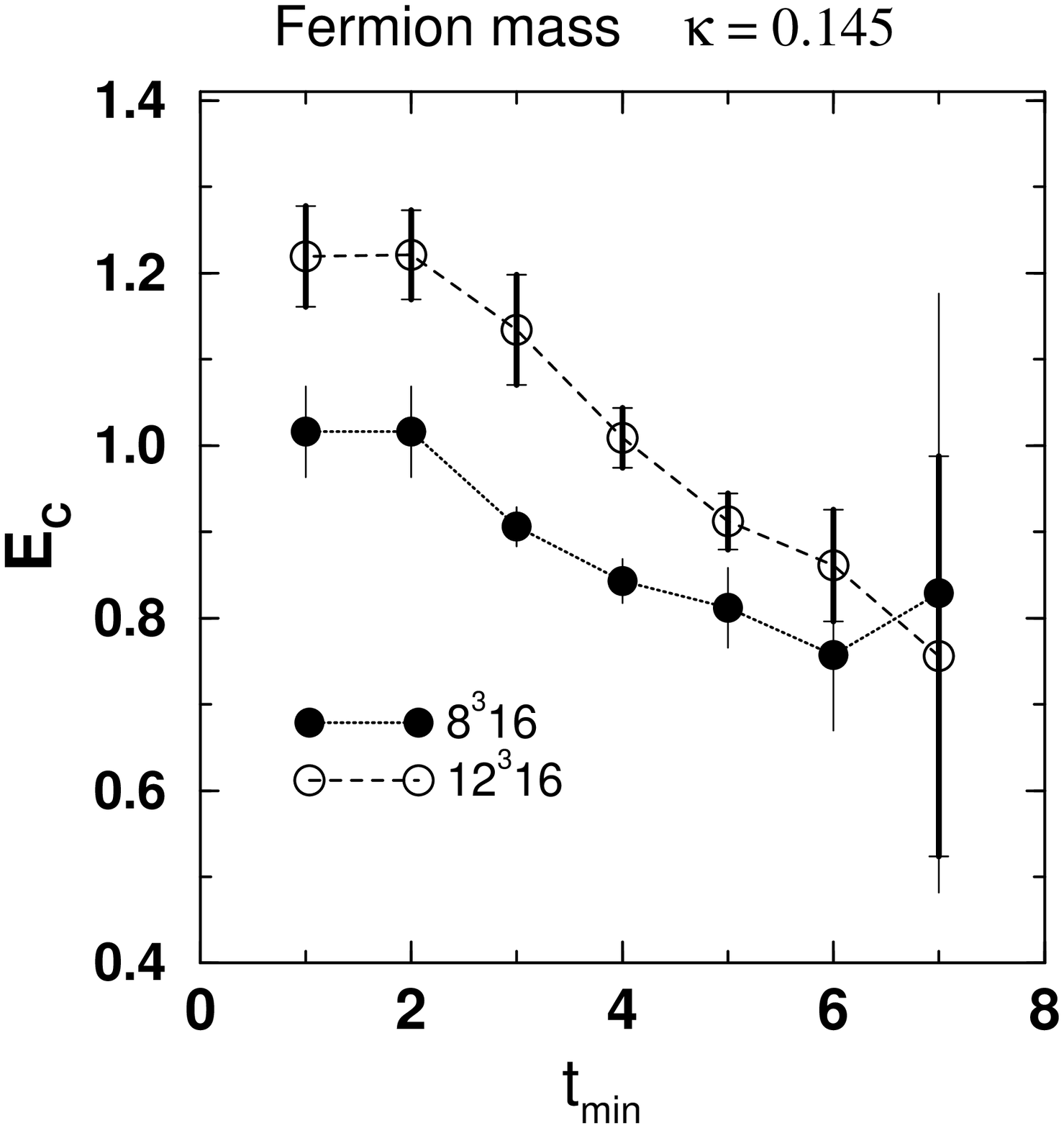}}
\end{picture}
\end{center}
\caption[The instability of the charged mass ]{ }
\label{figure:char-stabl}
\end{figure}
\begin{figure}[h]
\begin{center}
\begin{picture}(360,450)
\epsfxsize=12.0cm
\epsfysize=7.0cm
\put(0,190){\epsfbox{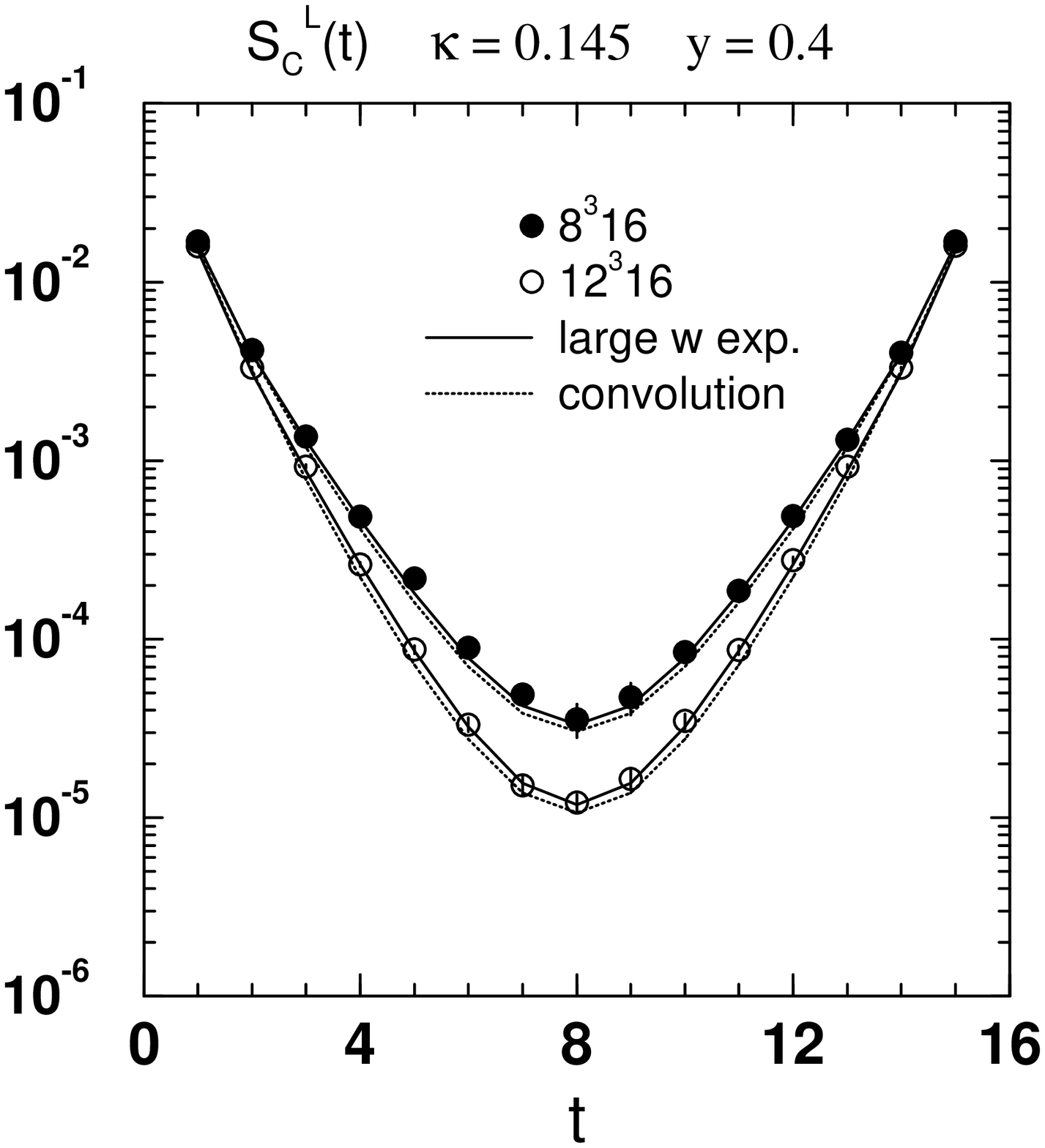}}
\end{picture}
\end{center}
\caption[The charged fermion propagator ($S^L$) by 1/w expansion
in the symmetric phase at $y=0.4$]{ }
\label{figure:char-1w}
\end{figure}
\begin{figure}[h]
\begin{center}
\begin{picture}(360,450)
\epsfxsize=12.0cm
\epsfysize=7.0cm
\put(0,190){\epsfbox{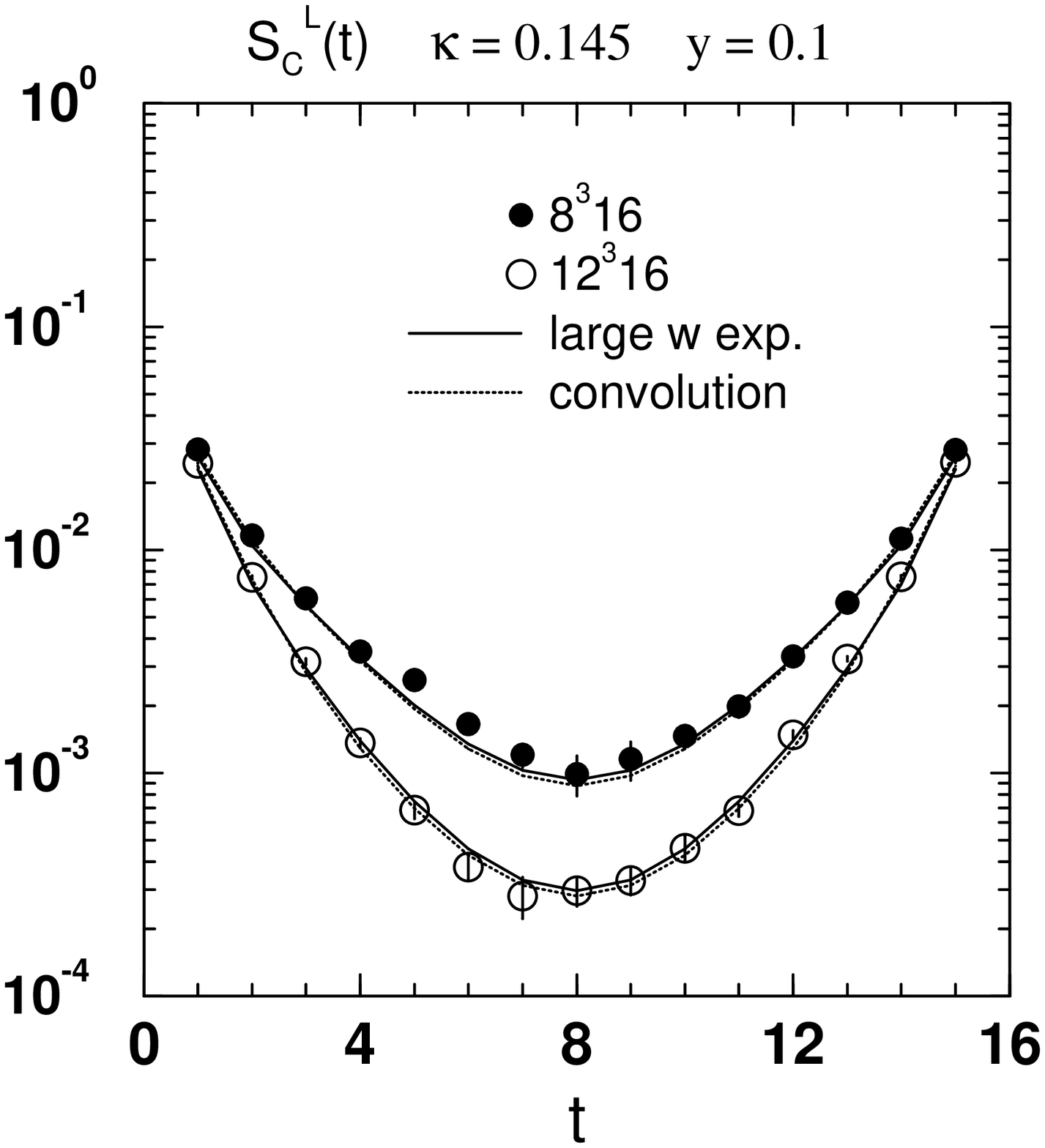}}
\end{picture}
\end{center}
\caption[The charged fermion propagator ($S^L$) 1/w expansion
in the symmetric phase at $y=0.1$]{ }
\label{figure:char-1w-0.1}
\end{figure}
\begin{figure}[h]
\begin{center}
\begin{picture}(360,450)
\epsfxsize=12.0cm
\epsfysize=7.0cm
\put(0,190){\epsfbox{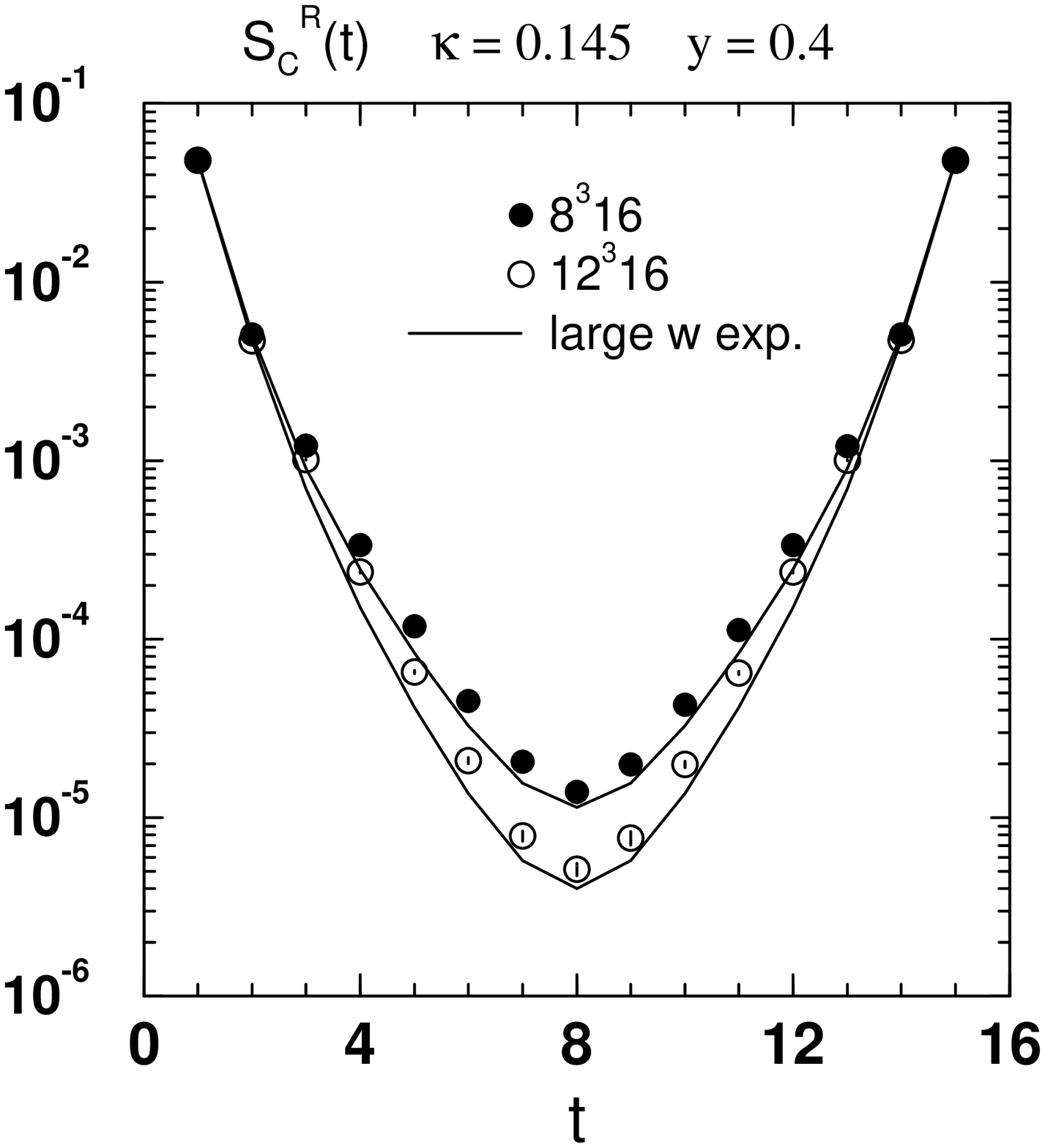}}
\end{picture}
\end{center}
\caption[The charged propagator $S^R$ by 1/w expansion at y = 0.4]{ }
\label{figure:char-1w-SR}
\end{figure}
\begin{figure}[h]
\begin{center}
\begin{picture}(360,450)
\epsfxsize=12.0cm
\epsfysize=7.0cm
\put(0,190){\epsfbox{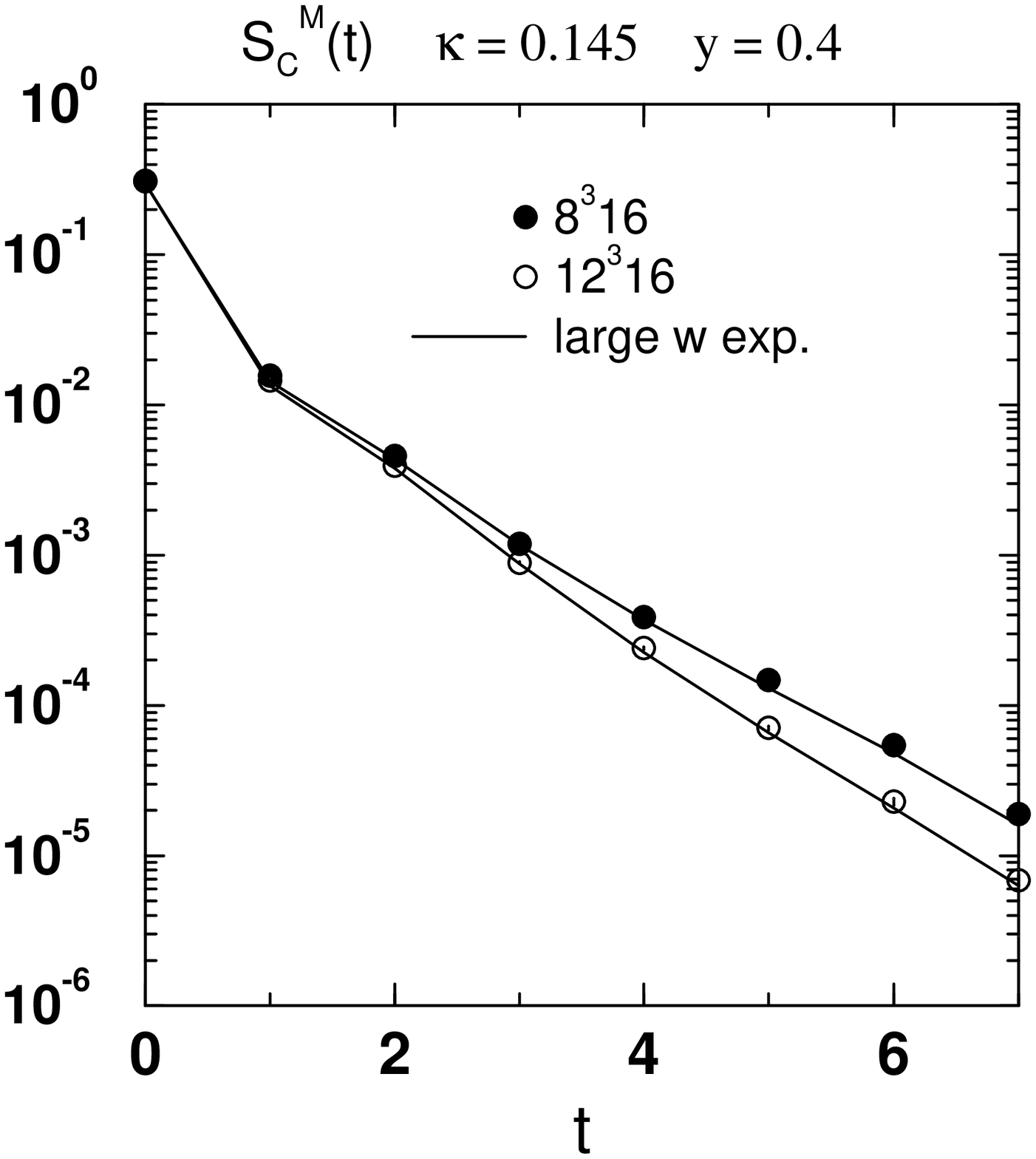}}
\end{picture}
\end{center}
\caption[The charged propagator $S^M$ by 1/w expansion at y = 0.4]{ }
\label{figure:char-1w-SM}
\end{figure}
\begin{figure}[h]
\begin{center}
\begin{picture}(360,450)
\epsfxsize=12.0cm
\epsfysize=7.0cm
\put(0,190){\epsfbox{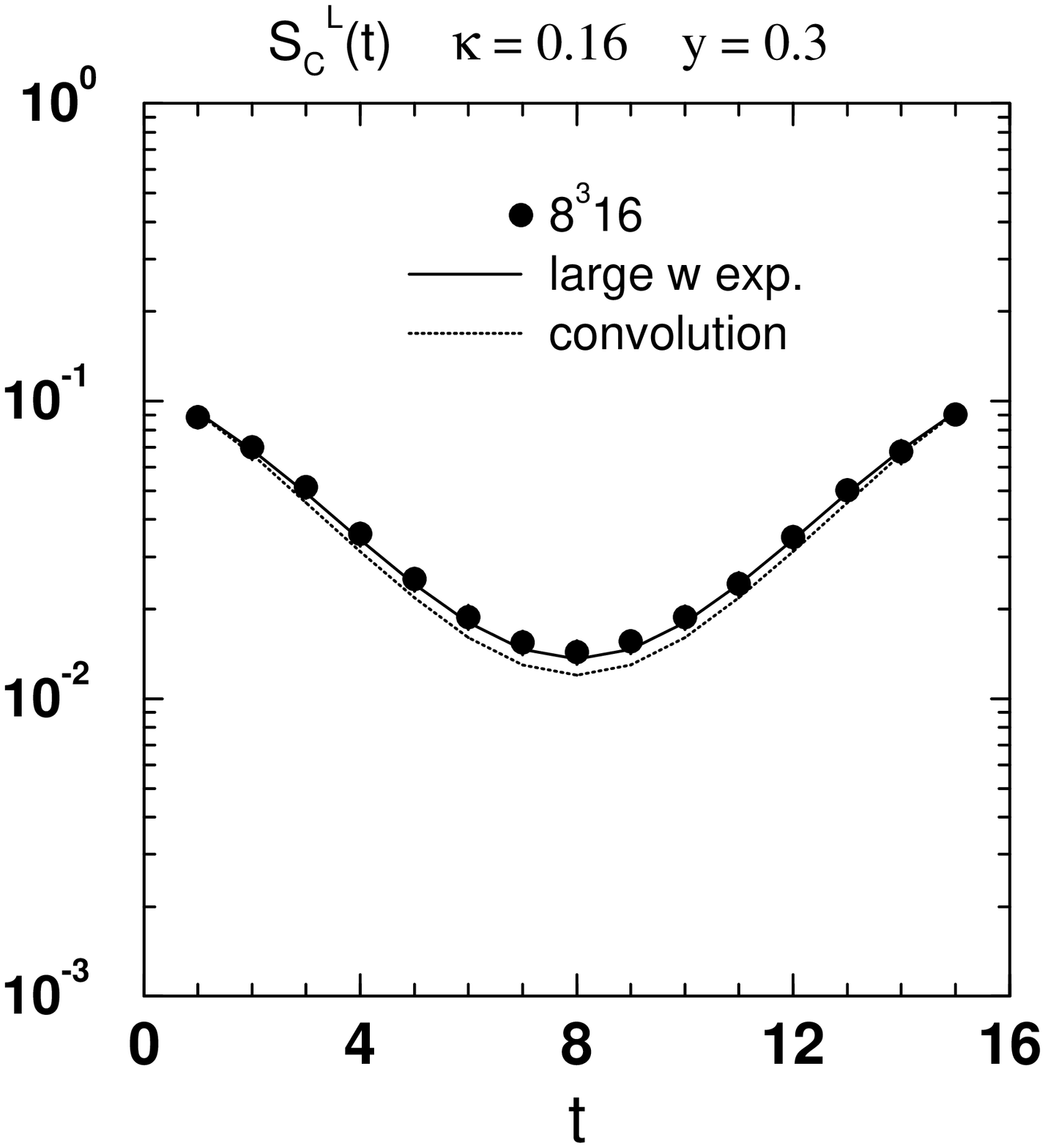}}
\end{picture}
\end{center}
\caption[The charged propagator $S^L$ by 1/w expansion
at y = 0.3 in the broken phase.]{ }
\label{figure:char-1w-bro}
\end{figure}
\end{document}